\documentclass{aa}

\usepackage{graphicx, subfigure}
\usepackage[varg]{txfonts}
\usepackage{natbib}
\usepackage{caption}
\bibpunct{(}{)}{;}{a}{}{,} 
\newcommand{\phn}   {\phantom{0}}

\newcommand{\phnnn} {\phantom{000}}
\begin{document} 

\title{The CO-H$_2$ van der Waals complex and complex organic molecules in cold molecular clouds: a TMC-1C survey\thanks{Based on observations carried out with the IRAM 30m Telescope. IRAM is supported by INSU/CNRS (France), MPG (Germany) and IGN (Spain).}}

  \author{A. Potapov\inst{1,2}
          \and
          \'A. S\'anchez-Monge\inst{1}
          \and
          P. Schilke\inst{1}
          \and
          U.~U. Graf\inst{1}
          \and
          Th. M\"oller\inst{1}
          \and
          S. Schlemmer\inst{1}
          }

  \institute{I.\ Physikalisches Institut, Universit\"at zu K\"oln, Z\"ulpicher Str.\ 77, 50937 Cologne, Germany
             \and
             Laboratory Astrophysics Group of the Max Planck Institute for Astronomy at the Friedrich Schiller University Jena, Institute of Solid State Physics, Helmholtzweg 3, 07743 Jena, Germany \\
             \email{alexey.potapov@uni-jena.de}}

  \date{Received ; accepted }

  \abstract{Almost 200 different species have been detected in the interstellar medium (ISM) during the last decades, revealing not only simple species but complex molecules with more than 6 atoms. Other exotic compounds, like the weakly-bound dimer (H$_2$)$_2$, have also been detected in astronomical sources like Jupiter.}
{We aim at detecting for the first time the CO-H$_2$ van der Waals complex in the ISM, which if detected can be a sensitive indicator for low temperatures.}
{We use the IRAM\,30m telescope, located in Pico Veleta (Spain), to search for the CO-H$_2$ complex in a cold, dense core in TMC-1C (with a temperature of $\sim 10$~K). All the brightest CO-H$_2$ transitions in the 3~mm (80-110~GHz) band have been observed with a spectral resolution of 0.5--0.7~km~s$^{-1}$, reaching a rms noise level of $\sim 2$~mK. The simultaneous observation of a broad frequency band, 16~GHz, has allowed us to conduct a serendipitous spectral line survey.}
{No lines belonging to the CO-H$_2$ complex have been detected. We have set up a new, more stringent upper limit for its abundance to be [CO-H$_2$]/[CO]~$\sim 5\times10^{-6}$, while we expect the abundance of the complex to be in the range $\sim 10^{-8}$--$10^{-3}$. The spectral line survey has allowed us to detect 75 lines associated with 41 different species (including isotopologues). We detect a number of complex organic species, e.g.\ methyl cyanide (CH$_3$CN), methanol (CH$_3$OH), propyne (CH$_3$CCH) and ketene (CH$_2$CO), associated with cold gas (excitation temperatures $\sim 7$~K), confirming the presence of these complex species not only in warm objects but also in cold regimes.}
{} 


   \maketitle

%
\section{Introduction\label{s:intro}}

During the last decades, the number of atomic and molecular species detected in the interstellar medium (ISM) has increased considerably thanks to (i) the improved sensitivity of facilities like the IRAM\,30m telescope in Spain or the Atacama Large Millimeter/submillimeter array (ALMA) in Chile, and (ii) new laboratory measurements of transitions of new species included in catalogues such as the Cologne Database for Molecular Spectroscopy (CDMS). Almost 200 different species have been found in Galactic/extragalactic environments such as cold dense cores, hot molecular cores, circumstellar disks, evolved stars or large diffuse molecular clouds. These 200 species do not consist only of simple molecules like the most abundant H$_2$ and CO, but also include complex species usually defined as molecules with 6 or more atoms (\citealt{HvD2009}; see the CDMS database\footnote{http://www.astro.uni-koeln.de/cdms/molecules} for a summary of detected species in space).

Molecular hydrogen, H$_2$, is by far the most abundant molecule in the Universe, followed by carbon monoxide, CO. Therefore, the intermolecular forces between these two species are of fundamental interest. If the CO-H$_2$ van der Waals complex\footnote{We note that this complex does not correspond to the formaldehyde molecule, H$_2$CO.} exists in measurable amounts in the ISM, it could be a sensitive indicator for low temperatures. The binding energy of the complex is so small --- typically 20~cm$^{-1}$ or 30~K --- that the relative abundance of the complex in the gas phase is expected to increase at lower temperatures.

There is an open debate about the feasibility of observing such weakly-bound species because their formation rates at the very low densities of interstellar molecular clouds (below 10$^{7}$~cm$^{-3}$) are low, due to the small probability of three-body collisions, which is the main formation route of van der Waals complexes in the laboratory. On the other hand, the large timescale on which these processes occur in interstellar space makes radiative association, which is usually a slow process, quite feasible (e.g.\ \citealt{Klem06}). Also non-equilibrium conditions in the ISM may strongly favor the formation and concentration of the CO-H$_2$ complex over time on the surfaces of dust grains in shielded regions at low temperatures, with release to the gas-phase occurring by localized heating processes such as turbulence or jets/outflows (e.g.\ \citealt{Allen97}). However, one also has to consider that CO tends to be frozen out onto dust grains in very cold, dense regions, and it seems difficult to release the CO-H$_2$ complex from grains without destroying it. On the other hand, this is completely unchartered territory, and even sensitive upper limits are useful. A detection of this complex would challenge many beliefs we have about the chemistry of dense molecular clouds.

There have been several attempts to observe complexes containing CO and H$_2$ molecules. The detection of the H$_2$ dimer, (H$_2$)$_2$, in the atmosphere of Jupiter has been reported by \citet{Mck88}, while searches for the CO dimer, (CO)$_2$ \citep{Van79}, and the CO-paraH$_2$ complex \citep{Allen97} in Galactic molecular clouds were not successful thus far. Laboratory data have clarified later a spectroscopic problem of these unsuccessful searches. The extensive millimeter-wave (MMW) studies of the CO dimer \citep{Su07} have shown that the radio astronomical search of this complex was based on frequencies which cannot be unambiguously attributed to (CO)$_2$. In the case of the CO-paraH$_2$ complex, the interstellar search was outside the correct frequency position of the most promising R(0) line, as later identified by the first MMW study of CO-paraH$_2$ \citep{Pak99}, and only the weaker Q(1) line was correctly tuned.

Recent laboratory studies of the CO-H$_2$ complex have provided precise MMW frequencies with uncertainties of about 50~kHz for the complex in different spin modifications and for its deuterated isotopologues: CO-paraH$_2$ \citep{Po09ApJ}, CO-orthoH$_2$ \citep{Jan13}, CO-orthoD$_2$ \citep{Po09OptSp} and CO-HD \citep{Po15}. Therefore, the availability of precise rest frequencies and modern astronomical receivers (with a sensitivity several times better than the old receivers used 20 years ago), combine in a great opportunity to detect for the first time a van der Waals complex in the ISM.

In this paper we present IRAM\,30m observations of a cold region in the Taurus molecular cloud in the search for the CO-H$_2$ van der Waals complex. In Sect.~\ref{s:obs} we describe the observations. In Sect.~\ref{s:res} we present the main results. Unfortunately, we do not have a detection of the CO-H$_2$ complex but we can set a new limit that can be used in future chemical modelling. In addition to the search for the CO-H$_2$ complex, the IRAM\,30m observations allowed us to conduct a spectral line survey of a very cold region ($\sim 10$~K), and we report the detection of complex organic molecules (COMs) as well as first time tentative detections of species in this object. In Sect.~\ref{s:disc} we discuss our results, and we end the paper with a summary of the main results in Sect.~\ref{s:summary}.

\begin{figure}[t]
\begin{center}
\begin{tabular}[b]{c}
    \includegraphics[width=0.9\columnwidth]{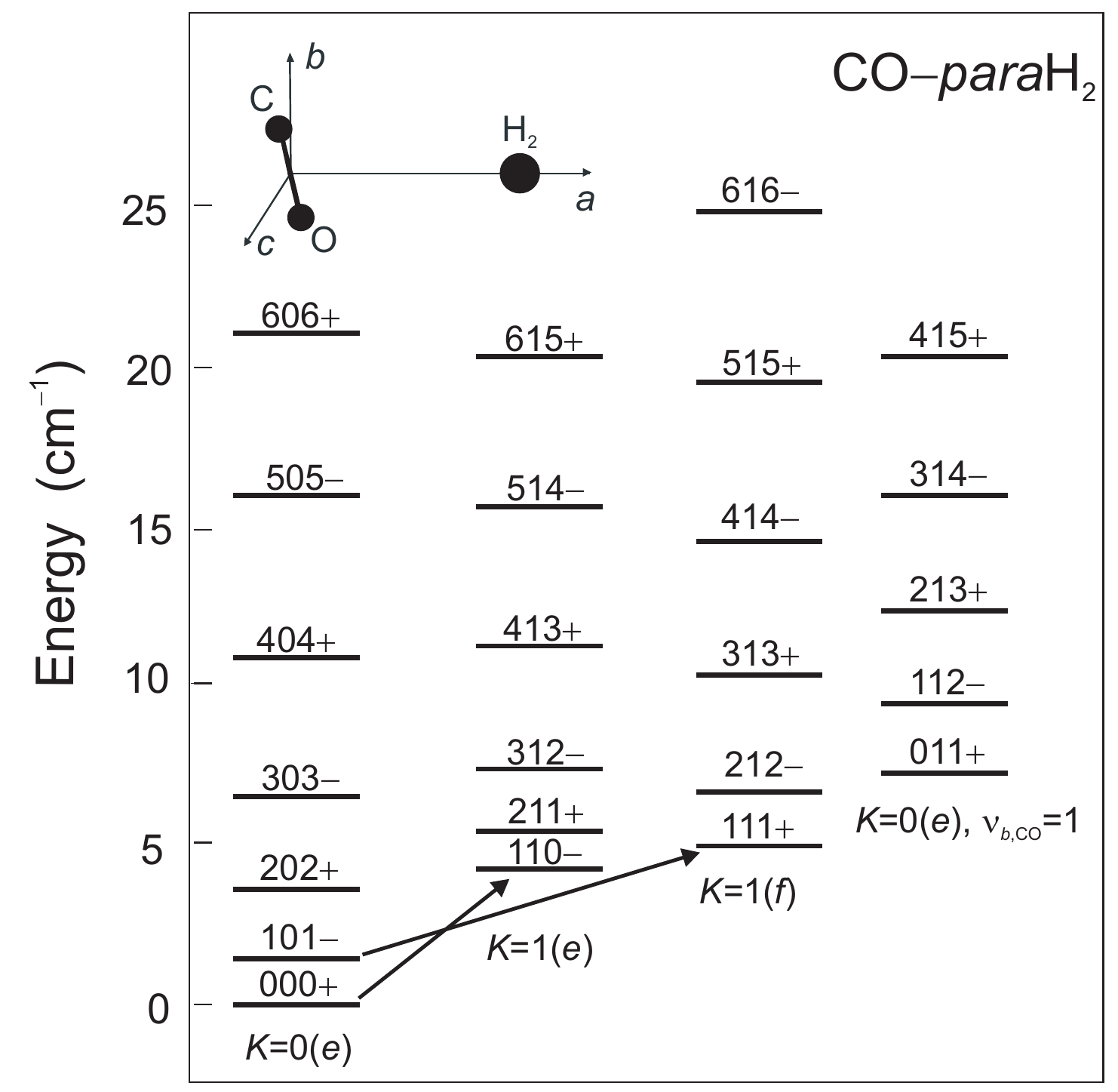} \\
    \includegraphics[width=0.9\columnwidth]{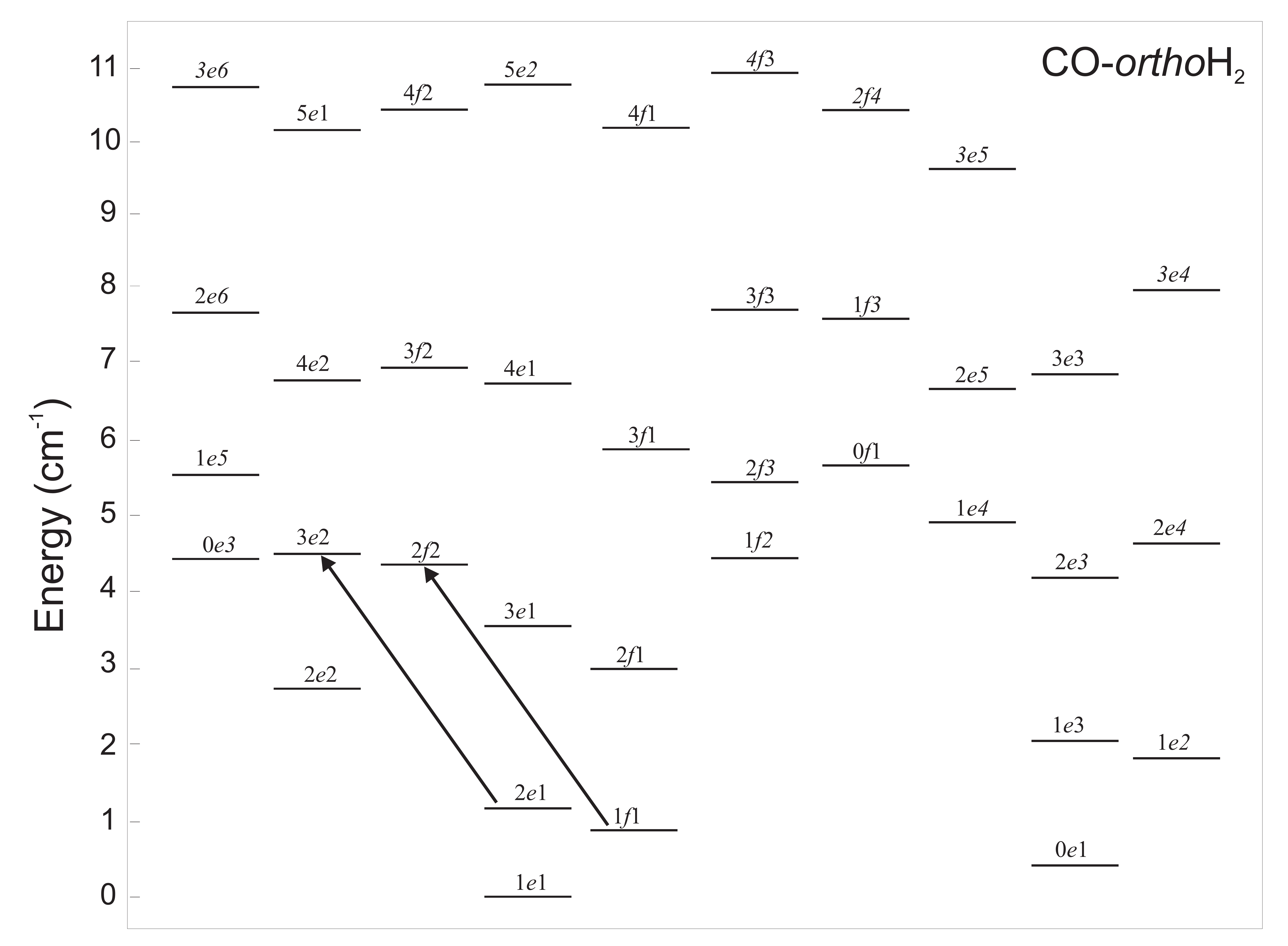} \\
\end{tabular}
\end{center}
\caption{Energy level diagram for the CO-paraH$_2$ (top panel) and CO-orthoH$_2$ (bottom panel) van der Waals complex. \textit{Top panel}: The energy levels are labeled by quantum numbers $J$, $j_\mathrm{CO}$ and $l$ where $J$ is the total angular momentum, $j_\mathrm{CO}$ refers to the rotation of the CO sub-unit and $l$ refers to the end-over-end rotation of the complex. $K$ corresponds to the projection of $J$ onto the intermolecular axis. The labels $e$ and $f$ indicate the parity of the $J$ levels within a given stack. The parity of an even-$J$ level is `$+$' for stacks labeled by $e$ and `$-$' for $f$, while the parity of an odd-$J$ level is `$-$' for $e$ and `$+$' for stacks labeled $f$. The insert shows the approximate geometrical structure of the CO-H$_2$ complex (see \citealt{Po09ApJ} for details). \textit{Bottom panel}: The energy levels are labeled by quantum numbers $J$, parity $P$ and $n_J$,$P$, a consecutive number of the state for the given values of $J$ and $P$ (see \citealt{Jan13} for details). In both panels, the targeted transitions are indicated by arrows.}
\label{f:energylevels}
\end{figure}

\begin{table}
\caption{Frequencies of the brightest CO-H$_2$ targeted lines}
\label{t:transitions}
\centering
\begin{tabular}{l c c}
\hline\hline

&Transition
&Frequency (MHz)
\\
\hline
CO-paraH$_2$  &(1,1,0)--(0,0,0) & 108480.857 \\
              &(1,1,1)--(1,0,1) & \phn91012.364 \\
CO-orthoH$_2$ &(2,f,2)--(1,f,1) & 101907.919 \\  
              &(3,e,2)--(2,e,1) & \phn93433.726 \\
CO-orthoD$_2$ &(1,1,0)--(0,0,0) & 102791.612 \\     
              &(1,1,1)--(1,0,1) & \phn89483.510 \\
CO-HD         &(1,1,0)--(0,0,0) & 105636.\phnnn \\
\hline
\end{tabular}
\tablefoot{The uncertainty in the frequency measurements is about 50~kHz for all the transitions, except for CO-HD with a few tens of MHz. The labelling of the quantum numbers of the transitions are explained in detail in Fig.~\ref{f:energylevels}.}
\end{table}

\begin{figure*}[h!]
\begin{center}
\begin{tabular}[b]{c}
        \includegraphics[width=0.95\textwidth]{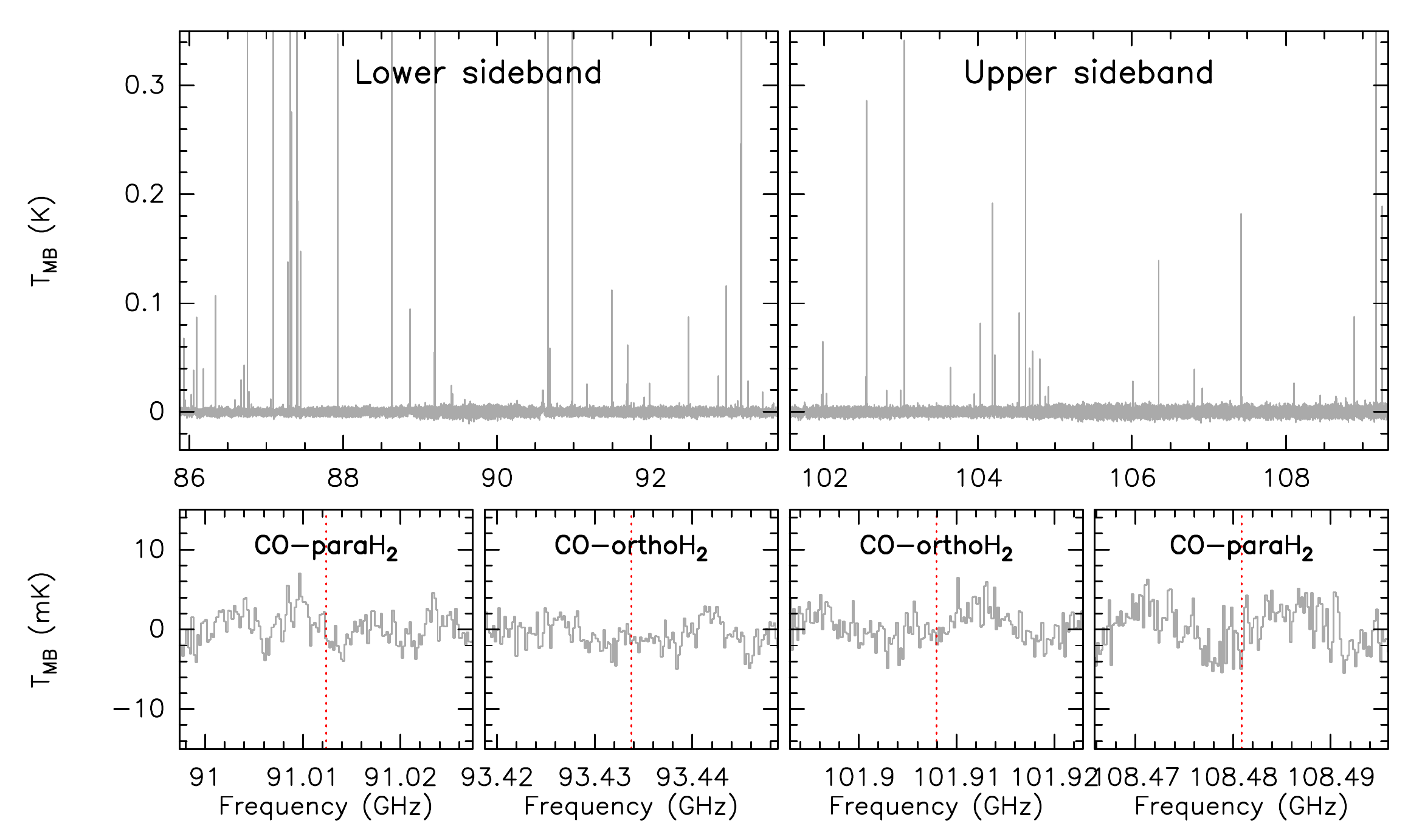} \\
\end{tabular}
\caption{\textit{Top panels}: Full spectrum observed with the IRAM\,30m telescope towards the cold dense core in TMC-1C. The mean rms noise level is $\sim 2$~mK. Most of the detected lines emit only in one channel (channel width 0.5--0.7~km~s$^{-1}$), suggesting that the linewidth of the different lines is $\le 0.7$~km~s$^{-1}$ (see Sect.~\ref{s:molecules}). \textit{Bottom panels}: Close-up views of the frequency ranges around the brightest transitions of the CO-H$_2$ van der Waals complex. The corresponding frequencies are listed in Table~\ref{t:transitions}, and are shown in the panels with a vertical dotted line. The expected linewidth is $\approx0.3$~km~s$^{-1}$, as measured in higher spectral resolution observations (e.g.\ \citealt{Spez13}).}
\label{f:specCOH2}
\end{center}
\end{figure*}

\section{Observations\label{s:obs}}

The observations were carried out from 2015 May 6 until May 9 at the IRAM\,30m telescope, located in Pico Veleta (Granada, Spain) under the project code 131-14. We have chosen to attempt the detection of the CO-H$_2$ complex in the nearest star forming region: the Taurus molecular cloud complex (e.g.\ \citealt{Ola88, Suz92, Rob00}), in particular towards a cold, dense condensation nearby TMC-1C which has measured low excitation temperatures of 3--7~K \citep{Spez13}, and for which a kinetic temperature of 10~K reproduce the observations presented by \citet{Spez16}. This object harbours the physical conditions (low temperature and high density, $\sim 4\times10^{4}$~cm$^{-3}$; \citealt{Schnee2007}) necessary to search for the CO-H$_2$ complex. We note that the density is still low enough to not have all the CO frozen out onto the dust grains\footnote{Referring to the work of \citet{Caselli1999}, a model in which CO is condensed out onto dust grains at densities above $10^5$~cm$^{-3}$ and has a roughly canonical abundance at lower hydrogen densities, is supported by the observations of gas-phase depletion in the L1544 cloud core.}. The coordinates used for the observations are $\alpha_{2000}$ = 04$^\mathrm{h}$41$^\mathrm{m}$16.$^\mathrm{s}$1 and $\delta_{2000}$ = +25\degr49\arcmin43\farcs8, coincident with the coordinates used in \citet{Spez13}.

We tuned the telescope to cover a number of transitions of the CO-H$_2$ complex and its deuterated isotopologues in the 3~mm band (E090) of the EMIR receiver. All four EMIR sub-bands were connected to the Fast Fourier Transform Spectrometers (FTS), with a spectral resolution of 200~kHz, which results in $\sim 0.5$--$0.7$~km~s$^{-1}$ at the corresponding frequencies. In Table~\ref{t:transitions}, we list the most intense transitions of the complex covered in our spectral setup. The frequency coverage was selected in order to optimize the simultaneous search of the strongest CO-paraH$_2$ and CO-orthoH$_2$ lines. The energy level diagrams for CO-paraH$_2$ and CO-orthoH$_2$ are shown in Fig.~\ref{f:energylevels}. In total, our observations cover an effective bandwidth of 16~GHz, ranging from 85.87 to 93.65~GHz in the lower sideband, and from 101.55 to 109.33~GHz in the upper sideband. The total observing time was 20~hours. We used the ON-OFF observation mode, with the reference position located at the offset (800\arcsec, 600\arcsec). The telescope pointing was checked every 1.5~hours and was found to be accurate to $\sim 5$\arcsec, i.e.\ only a fraction of the beam size of the telescope at these frequencies: $\sim 30$\arcsec. The weather conditions were stable during the observations with zenith opacities of 0.02--0.07 and system temperatures of 80--100~K. The observed spectra was calibrated following the standard procedures, and analyzed using the GILDAS\footnote{The GILDAS software package is developed by the IRAM and Observatoire de Grenoble, and can be downloaded at http://www.iram.fr/IRAMFR/GILDAS} software package. We converted the spectra to the main beam temperature scale, using a forward efficiency of 0.95, and a beam efficiency of 0.79 and 0.81 for the upper and the lower sidebands, respectively. The final spectrum has a noise level of $\sim 2$~mK.

\section{Results and analysis\label{s:res}}

\subsection{The CO-H$_2$ complex\label{s:coh2}}

In Fig.~\ref{f:specCOH2} we show, in the top panels, the full spectrum obtained with the IRAM\,30m telescope. A number of bright lines have been detected throughout the covered frequency range and they will be discussed in Sect.~\ref{s:molecules}. The bottom panels of Fig.~\ref{f:specCOH2} show a close-up view of the frequency ranges around the frequencies of the brightest CO-H$_2$ lines, corresponding to the CO-orthoH$_2$ and CO-paraH$_2$ transitions listed in Table~\ref{t:transitions}. No lines belonging to the CO-H$_2$ complex are detected at the corresponding frequencies (indicated in the figure with red dotted vertical lines). It is important to note that the noise at the high-frequency transitions is slightly larger than the average one (i.e.\ 2~mK). This larger noise is due to ripples in the baseline that were not possible to completely remove. However, since their wavelength is much larger than the expected linewidths, $\sim 0.3$~km~s$^{-1}$, they do not affect the search for the CO-H$_2$ transitions. None of the four transitions have been detected, and we therefore, set an upper limit of $\sim 6$~mK (corresponding to 3$\sigma$) for the intensities of these lines.

\begin{table}
\caption{Species, temperatures, column densities and transitions detected above $5\sigma$ towards TMC-1C (see Sect.~\ref{s:molecules} for details)}
\label{t:molecules}
\centering
\begin{tabular}{l c c c}
\hline\hline
Species
&$T$ (K)
&log($N$) (log[cm$^{-2}$])
&Transitions
\\
\hline
$^{13}$C$^{18}$O	&7.0			&$13.30\pm0.50$	&1	\\
$^{13}$CS			&7.0			&$11.65\pm0.26$	&1	\\
SO					&$6.2\pm0.7$	&$13.25\pm0.28$	&2	\\
S$^{18}$O			&6.2			&$11.81\pm0.26$	&1	\\
CCH					&7.0			&$13.42\pm0.11$	&6	\\
CCS					&$4.7\pm0.7$	&$12.95\pm0.38$	&4	\\
HCN					&7.0			&$12.26\pm0.18$	&1	\\
H$^{13}$CN			&7.0			&$11.51\pm0.29$	&1	\\
HC$^{15}$N			&7.0			&$10.71\pm0.21$	&1	\\
HNC					&7.0			&$12.92\pm1.01$	&1	\\
HN$^{13}$C			&7.0			&$11.76\pm0.13$	&1	\\
H$^{15}$NC			&7.0			&$11.51\pm0.30$	&1	\\
HCO					&$6.1\pm1.8$	&$11.03\pm0.30$	&4	\\
HCO$^{+}$			&7.0			&$12.09\pm0.35$	&1	\\
H$^{13}$CO$^{+}$	&7.0			&$11.59\pm0.38$	&1	\\
HC$^{17}$O$^+$		&7.0			&\phn$9.97\pm1.50$	&1	\\
DCS$^{+}$			&7.0			&$10.88\pm0.24$	&1	\\
SO$_2$				&7.0			&$11.95\pm0.23$	&1	\\
N$_2$H$^+$			&7.0			&$12.42\pm0.19$	&1	\\
CCCS				&$6.2\pm0.5$	&$12.67\pm0.31$	&2	\\
CCCO				&7.0			&$11.26\pm0.23$	&1	\\
c-CCCH				&$6.0\pm1.9$	&$12.14\pm0.35$	&6	\\
H$_2$CS				&$14.8\pm2.1$	&$12.67\pm0.15$	&2	\\
HDCS				&14.8			&$12.01\pm0.98$	&1	\\
H$_2$C$^{34}$S		&14.8			&$11.43\pm1.59$	&1	\\
HOCO$^+$			&7.0			&$11.25\pm0.19$	&1	\\
HNCO				&7.0			&$13.11\pm0.11$	&1	\\
NH$_2$D				&7.0			&$12.59\pm1.00$	&1	\\
C$_4$H				&$7.8\pm0.9$	&$13.12\pm0.26$	&2	\\
l-C$_3$H$_2$		&$7.2\pm0.7$	&$11.18\pm0.31$	&3	\\
c-C$_3$HD			&$6.0\pm1.8$	&$12.14\pm0.35$	&4	\\
c-C$_3$D$_2$		&6.0			&$10.99\pm1.90$	&1	\\
CH$_2$CO			&7.0			&$12.42\pm0.37$	&1	\\
HC$_3$N				&$4.2\pm0.3$	&$14.62\pm0.45$	&2	\\
HCCNC				&7.0			&$11.81\pm0.15$	&1	\\
HCOOH				&7.0			&$11.64\pm0.13$	&1	\\
CH$_3$CN			&$7.1\pm1.0$	&$11.80\pm0.45$	&2	\\
CH$_3$OH			&7.0			&$12.86\pm1.27$	&1	\\
CH$_2$DOH			&7.0			&$11.60\pm0.50$	&1	\\
CH$_3$CCH			&$15.0\pm3.6$	&$13.31\pm0.13$	&3	\\
CH$_3$CCD			&15.0			&$11.97\pm0.16$	&2	\\
\hline
\end{tabular}
\end{table}

\begin{figure*}[t]
\begin{center}
\begin{tabular}[b]{c}
    \includegraphics[width=0.8\textwidth]{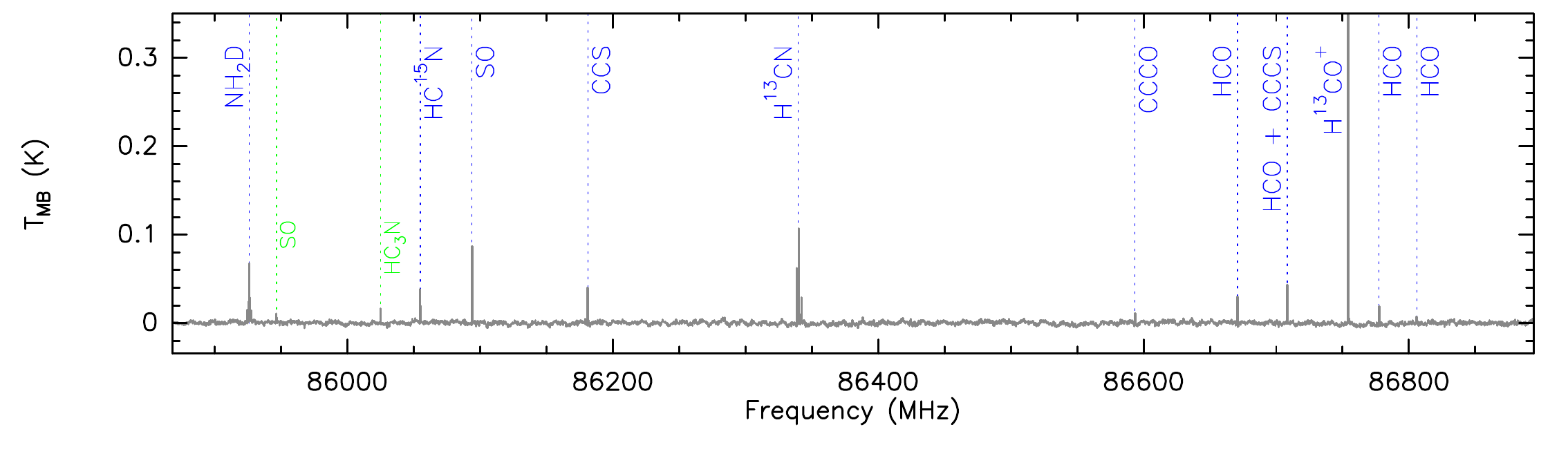} \\
    \includegraphics[width=0.8\textwidth]{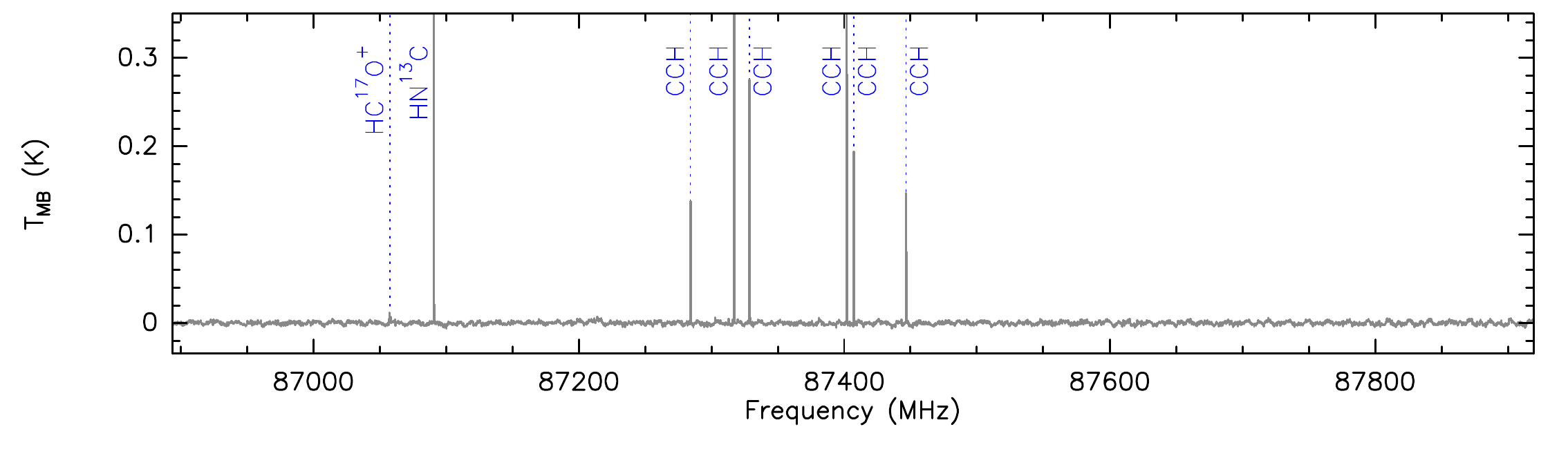} \\
    \includegraphics[width=0.8\textwidth]{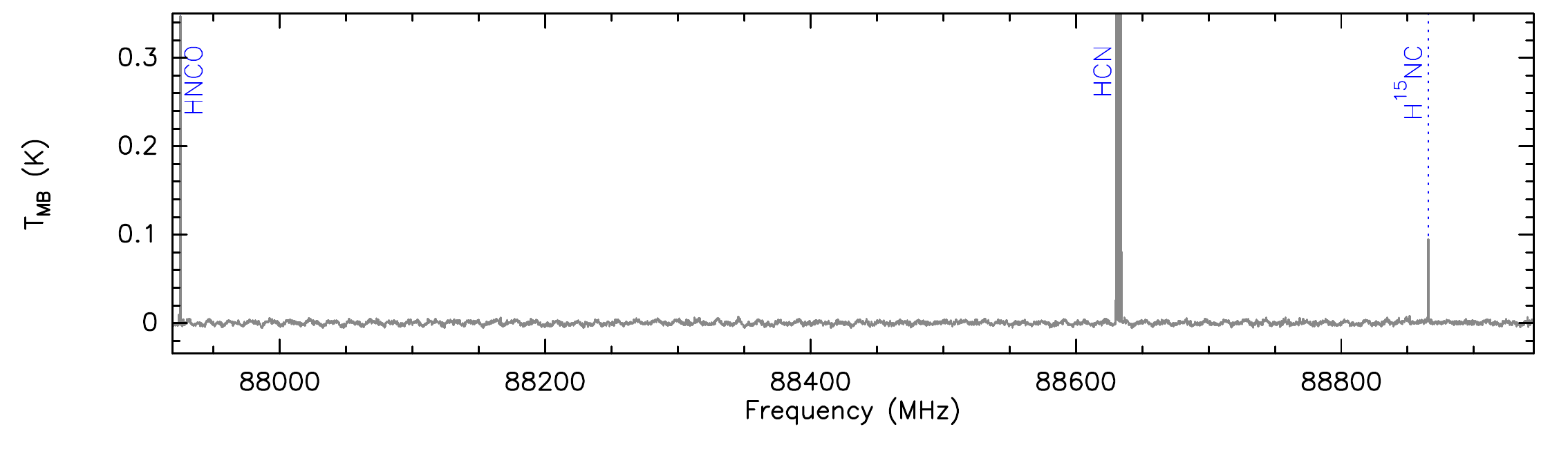} \\
    \includegraphics[width=0.8\textwidth]{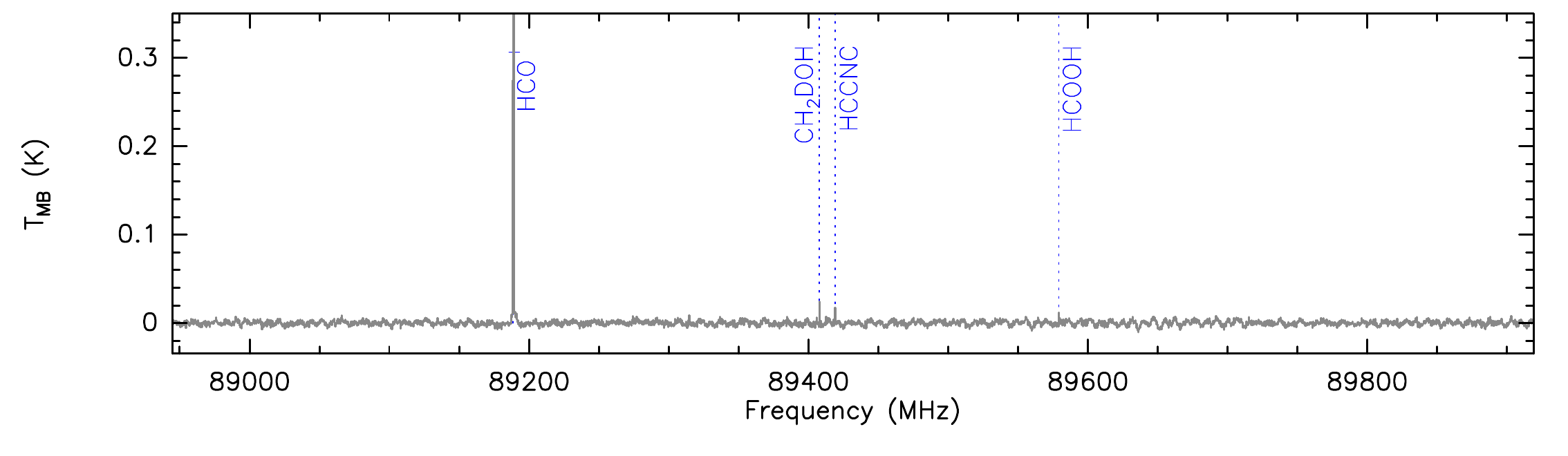} \\
\end{tabular}
\end{center}
\caption{Spectral line survey towards TMC-1C. Each panel shows about 1~GHz of the total 16~GHz frequency band. The observed spectrum is shown in dark grey. Each identified transition is indicated with a blue dotted line and the name of the corresponding species. The green dotted lines correspond to ghost lines from the image sideband.}
\label{f:molecules}
\end{figure*}
\begin{figure*}[t]
\ContinuedFloat
\begin{center}
\begin{tabular}[b]{c}
    \includegraphics[width=0.8\textwidth]{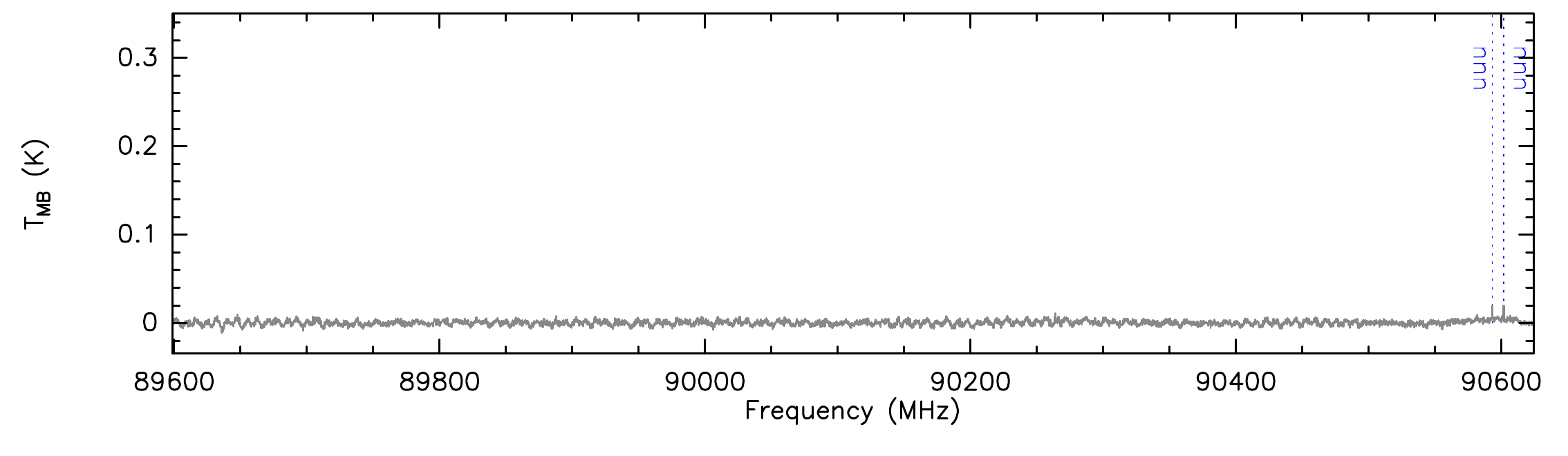} \\
    \includegraphics[width=0.8\textwidth]{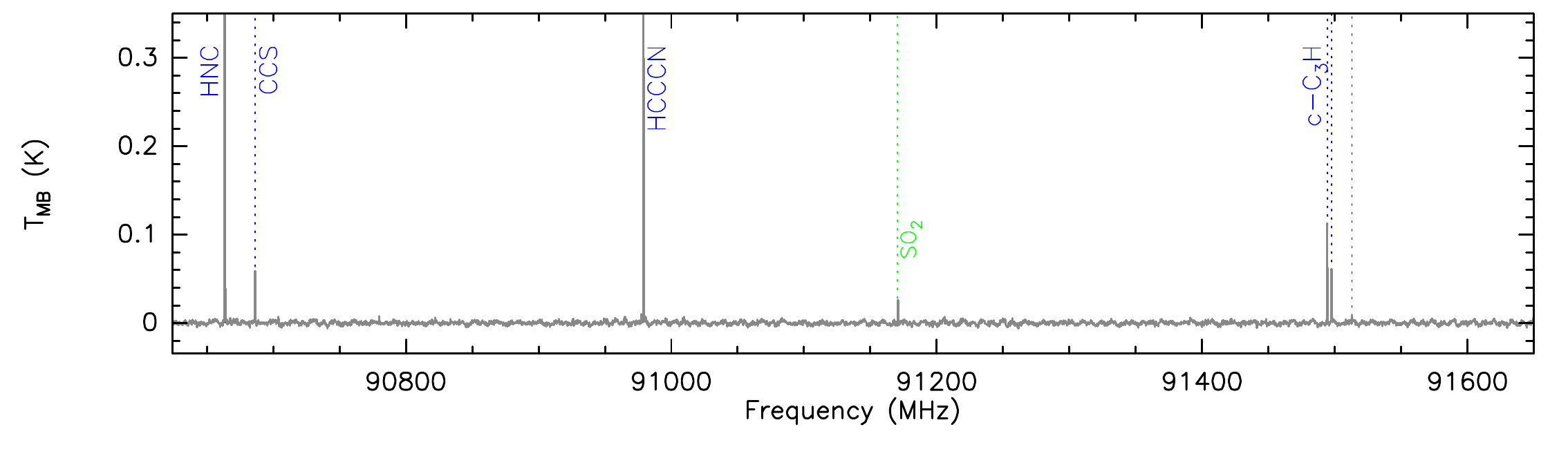} \\
    \includegraphics[width=0.8\textwidth]{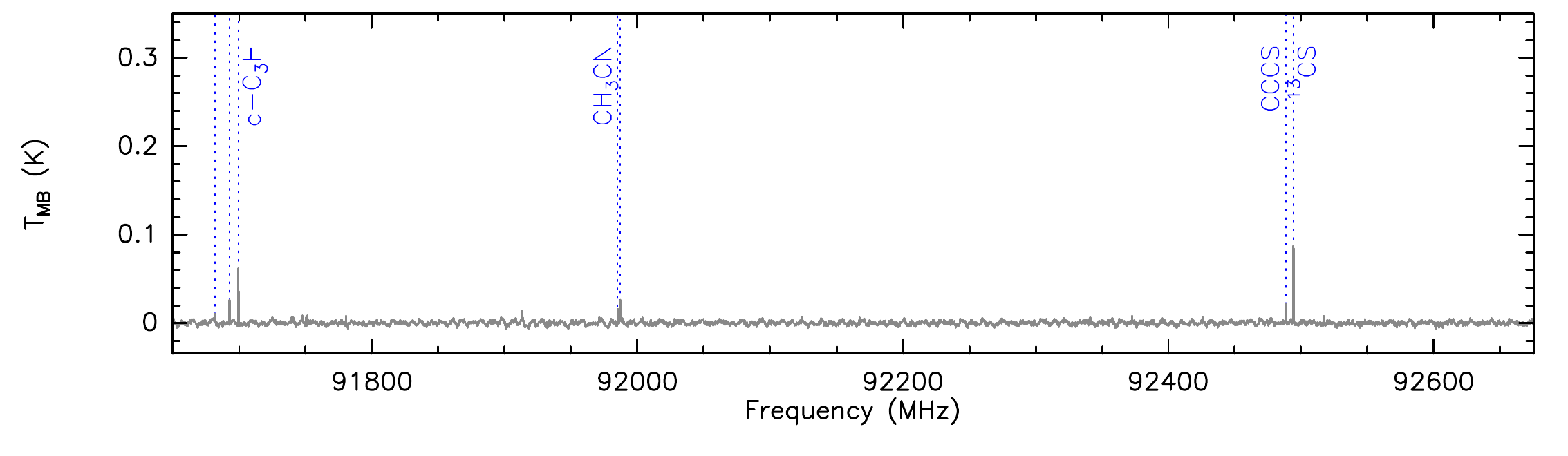} \\
    \includegraphics[width=0.8\textwidth]{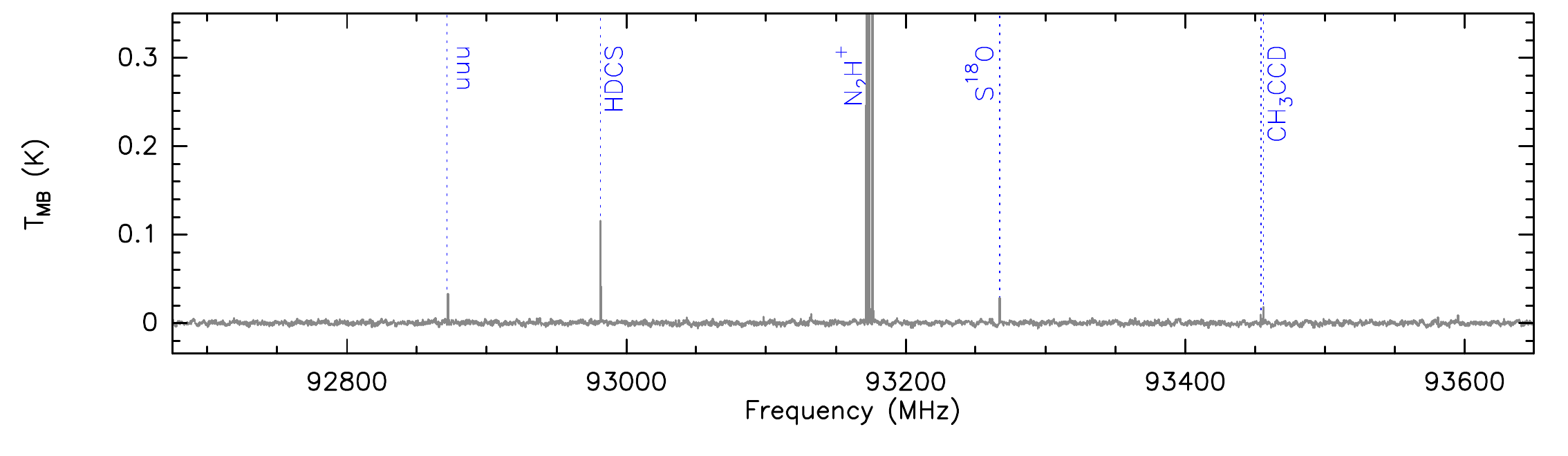} \\
\end{tabular}
\end{center}
\caption{Continued.}
\end{figure*}
\begin{figure*}[t]
\ContinuedFloat
\begin{center}
\begin{tabular}[b]{c}
    \includegraphics[width=0.8\textwidth]{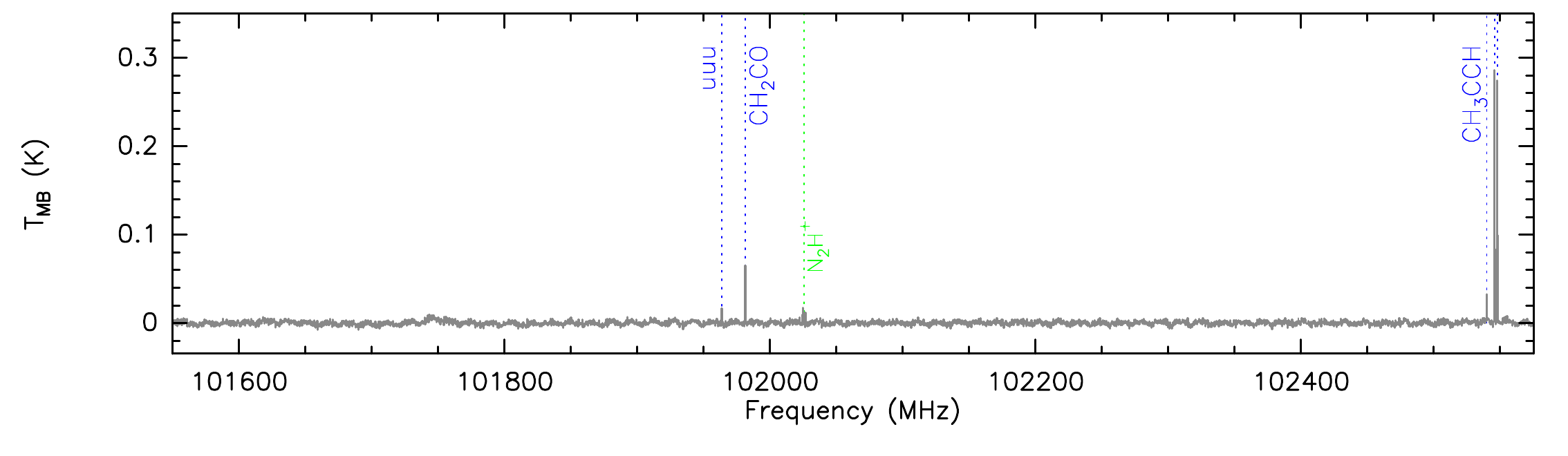} \\
    \includegraphics[width=0.8\textwidth]{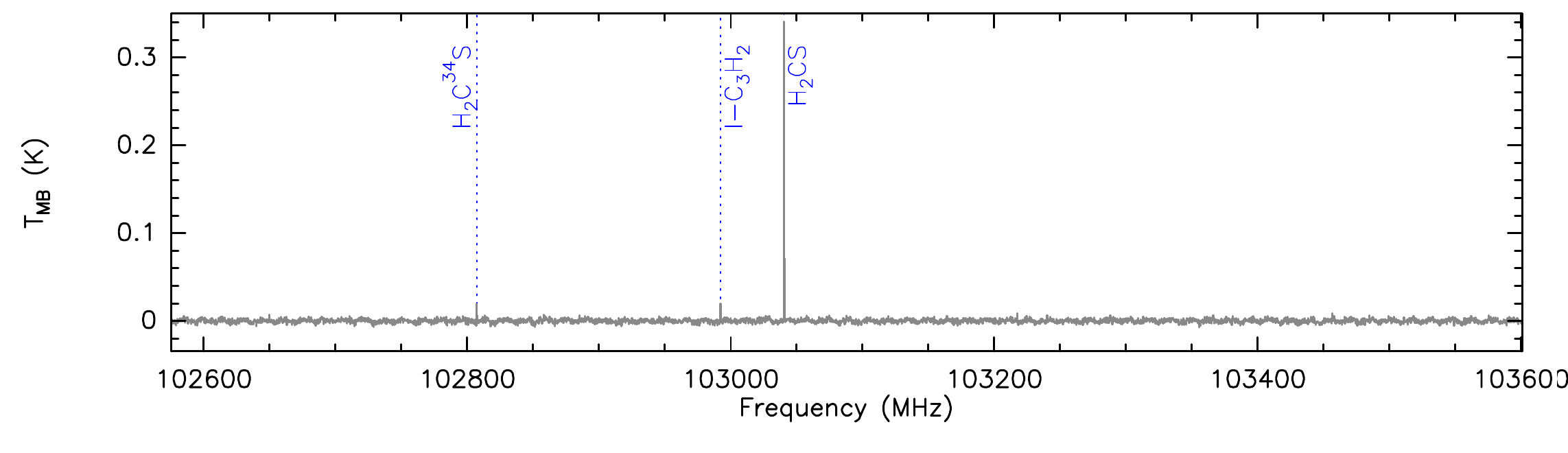} \\
    \includegraphics[width=0.8\textwidth]{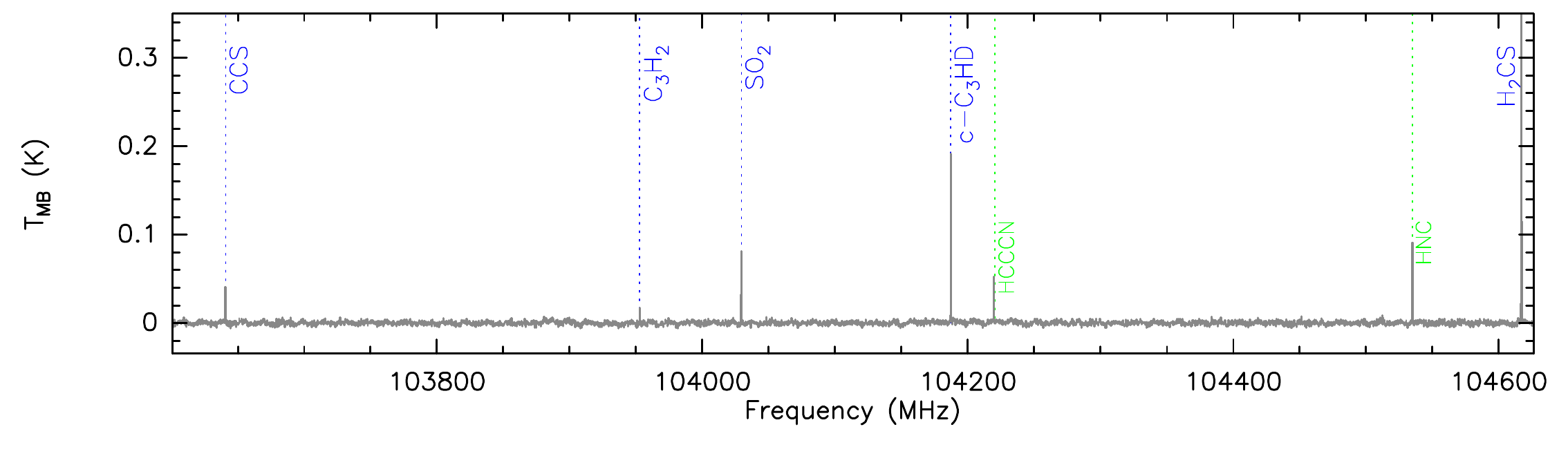} \\
    \includegraphics[width=0.8\textwidth]{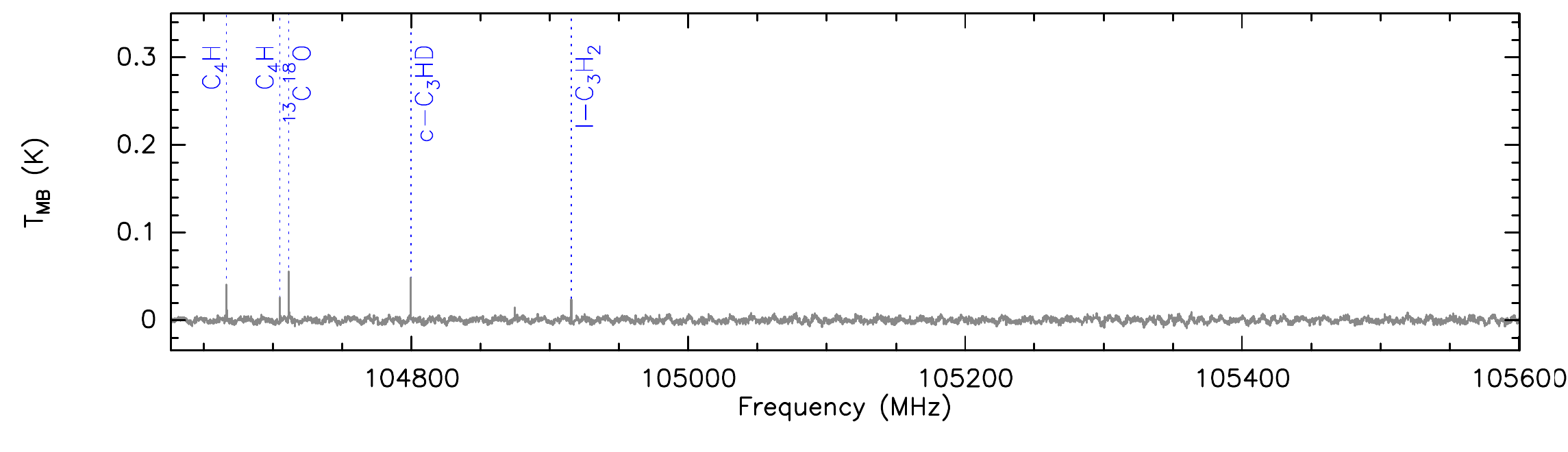} \\
\end{tabular}
\end{center}
\caption{Continued.}
\end{figure*}
\begin{figure*}[t]
\ContinuedFloat
\begin{center}
\begin{tabular}[b]{c}
    \includegraphics[width=0.8\textwidth]{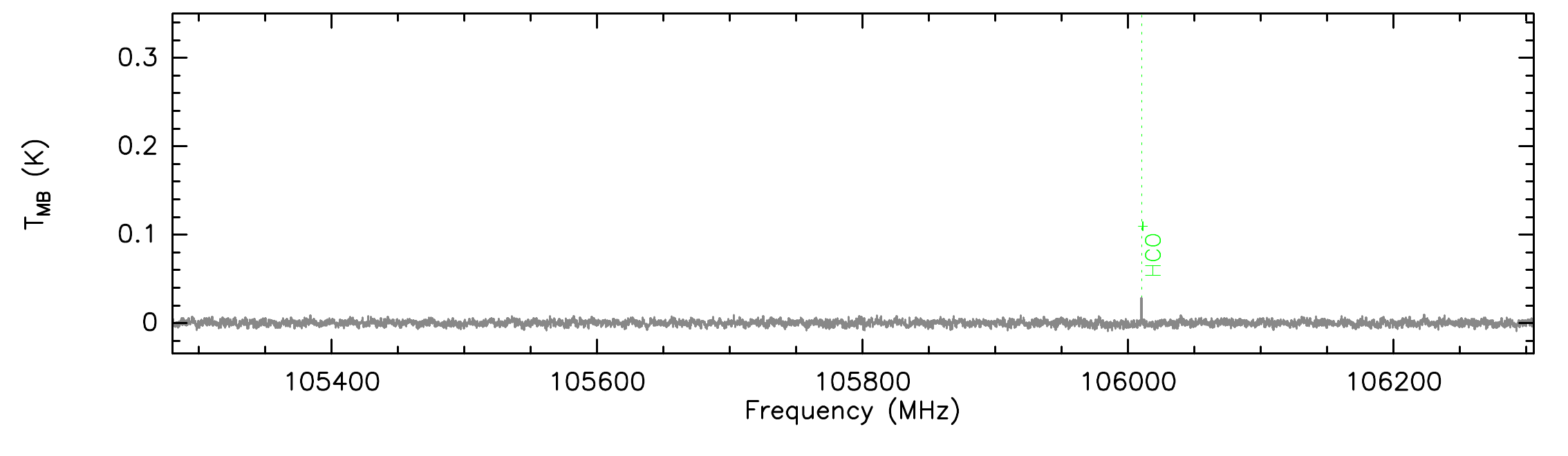} \\
    \includegraphics[width=0.8\textwidth]{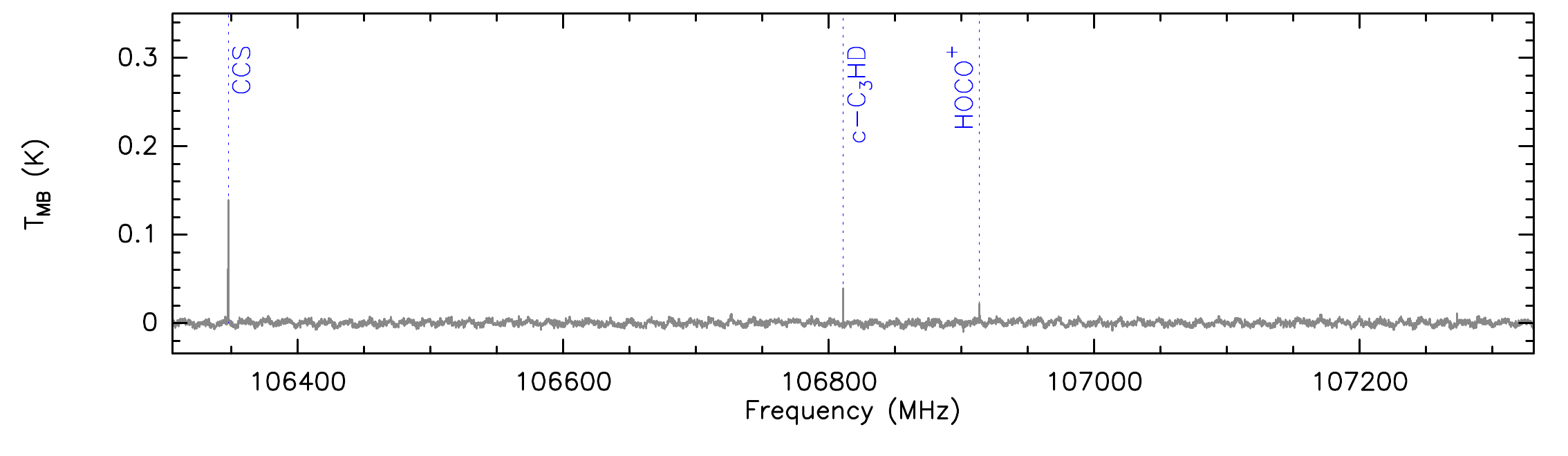} \\
    \includegraphics[width=0.8\textwidth]{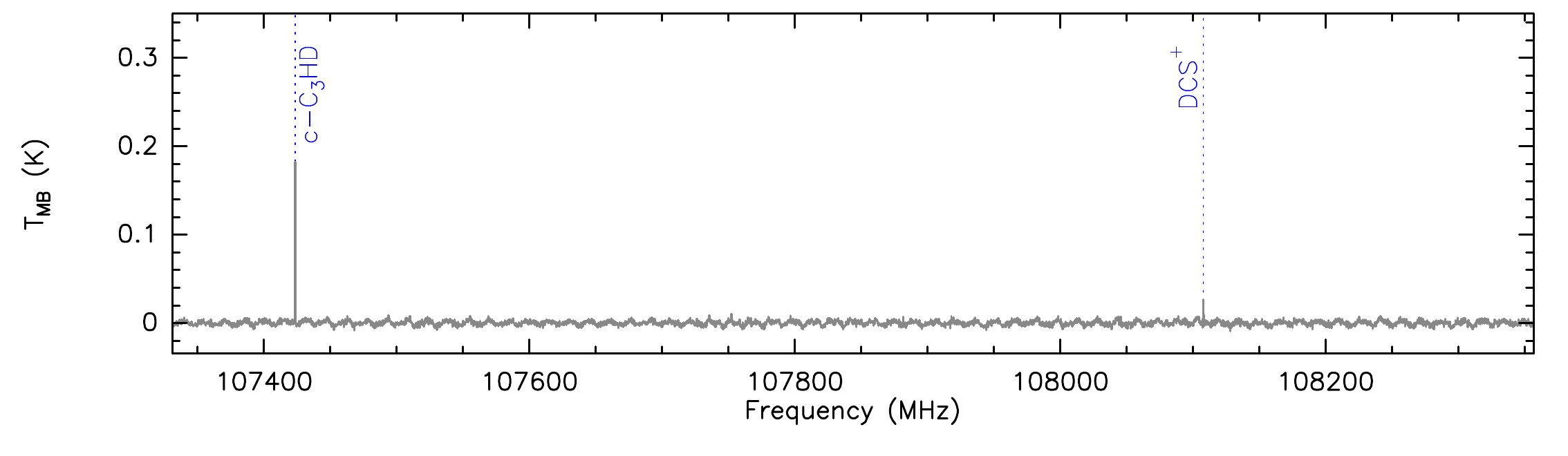} \\
    \includegraphics[width=0.8\textwidth]{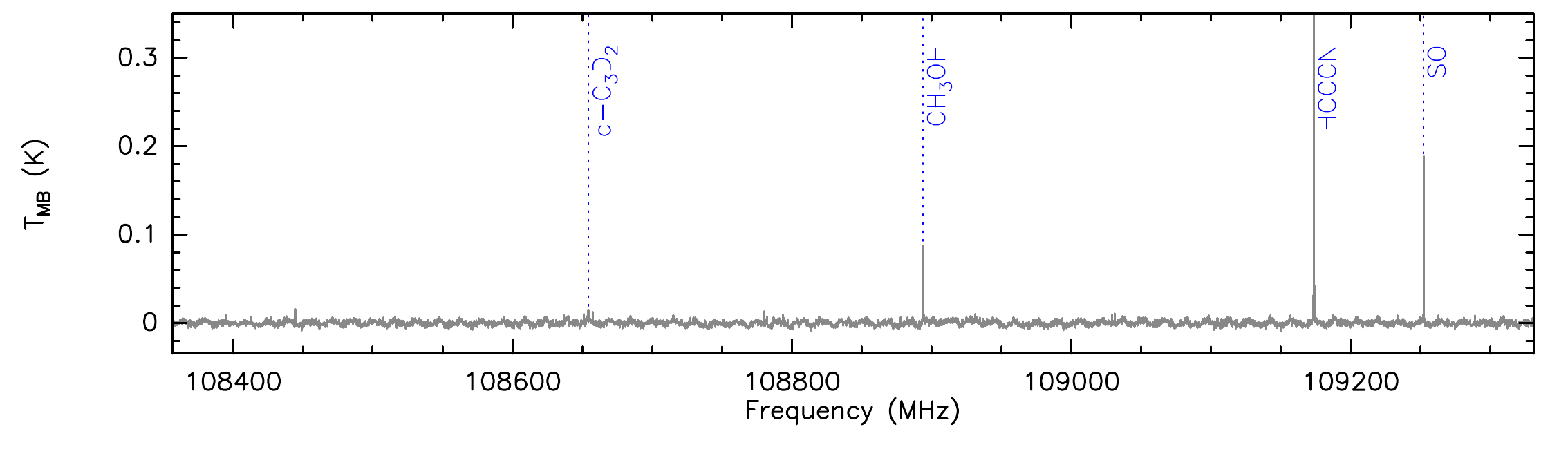} \\
\end{tabular}
\end{center}
\caption{Continued.}
\end{figure*}

\subsection{Spectral line survey\label{s:molecules}}

The broad frequency range covered with the IRAM\,30m telescope permits not only to study the CO-H$_2$ lines, but also to perform a spectral line survey of this cold dense condensation. It is worth mentioning that the high sensitivity achieved with our observations is adequate to search, e.g., for COMs (molecules with 6 or more atoms) in low-temperature environments. COMs have long been detected in the interstellar medium, especially in hot molecular cores associated with high-mass star forming regions (e.g.\ \citealt{Cum86, Bl87, SanchezMonge2013, SanchezMonge2014}). The advent of sensitive instruments has also revealed a chemical complexity associated with low-mass hot cores (or hot corinos; e.g.\ \citealt{Caz03}) and intermediate-mass hot cores (e.g.\ \citealt{SanchezMonge2010}). Despite their formation routes remain uncertain, both warm gas-phase and grain-surface reactions have been invoked to account for their presence in low-mass protostars. In this latter scheme, COMs result from radical-radical reactions on the grains as radicals become mobile when the nascent protostar warms up its surroundings and the resulting molecules are subsequently desorbed into the gas phase at higher temperatures or by shock events produced by winds/jets (e.g.\ \citealt{GnH06}). In the last years, the detection of COMs in cold environments ($T<30$~K; \citealt{Bac12, Vas14}) represents a challenge for the chemical models and an opportunity to improve and clarify the role of the grain-surface and gas-phase chemistry.

The top panels of Fig.~\ref{f:specCOH2} show the spectral line survey towards the condesantion in TMC-1C. We have identified 75 lines with an intensity $>5\sigma$. We note that the spectral resolution of only 0.5--0.7~km~s$^{-1}$ is not enough to resolve most of the lines, suggesting that all they are excited in an environment with a temperature $<30$~K, i.e\ in cold gas\footnote{Most of the lines are detected in one single channel. The exceptions are species such as HCN and N$_2$H$^+$ due to the hyperfine structure, and $^{13}$C$^{18}$O with a weak blue-shifted wing. The thermal linewidth for gas at 30~K is 0.23~km~s$^{-1}$, 0.18~km~s$^{-1}$ and 0.16~km~s$^{-1}$, for species with mean molecular weights of 25 (e.g.\ CCH), 40 (e.g.\ CH$_3$CCH) and 56 (e.g.\ CCS), respectively.}. The identification of the lines was done using the CDMS \citep{Mul05} and JPL \citep{Pic98} databases, and later on confirmed by creating synthetic spectra of each species using the XCLASS\footnote{The eXtended CASA Line Astronomy Software Suite (XCLASS) can be downloaded at https://www.astro.uni-koeln.de/projects/schilke/XCLASSInterface} software \citep{Moeller2015}. XCLASS is a toolbox for the Common Astronomy Software Applications (CASA) package containing new functions for modeling interferometric and single-dish data. The included myXCLASS program calculates synthetic spectra by solving the radiative transfer equation for an isothermal object in one dimension where the required molecular data are taken from an embedded database containing entries from the CDMS and JPL databases using the Virtual Atomic and Molecular Data Center (VAMDC) portal. The contribution of a molecule is defined by an user-defined number of components where each component is described by four main parameters: excitation temperature, column density, velocity width and velocity offset. In order to achieve a good description of the data we fit these parameters using the included model optimizer package MAGIX \citep{Moeller2013}. By performing an internal resampling, XCLASS makes sure that the line is sampled properly, even if the velocity resolution of the data is coarse.

In Table~\ref{t:molecules} we list the identified species with the corresponding excitation temperature (column~2), column density (column~3) and number of transitions above $5\sigma$ (column~4). The 75 detected lines come from 41 different species (including isotopologues) and four unidentified lines at the frequencies 90.593~GHz, 90.602~GHz, 92.872~GHz and 101.981~GHz with main beam temperatures of 18~mK, 18~mK, 30~mK and 15~mK, respectively. For those species with more than one transition we have fitted with XCLASS simultaneously the excitation temperature and column density\footnote{For species like C$_3$H$_2$ or CH$_3$OH, with different spin symmetries: ortho/para, the total column density is given.}. If only one transition is detected above $5\sigma$ for a given species, we fit only the column density and fix the excitation temperature to 7~K, which corresponds to the average temperature of the other transitions and is consistent with the value measured also in \citet{Spez13}. For the different detected isotopologues, we have fixed the excitation temperature to be the same of the main species, and we have fitted only the column density. In all cases we consider a linewidth of 0.3~km~s$^{-1}$. In Fig.~\ref{f:molecules} we present the whole spectral survey, indicating the location of the different detected transitions. A number of COMs have been detected in this cold dense core. We discuss the results in Sect.~\ref{s:moleculesDisc}.

\section{Discussion\label{s:disc}}

\subsection{Detectability of the CO-H$_2$ complex\label{s:coh2Disc}}

In Sect.~\ref{s:coh2} we report an upper limit of $\sim 6$~mK for the CO-H$_2$ lines. We follow the same approach as in \citet{Allen97} to establish an upper limit for the column density of the CO-paraH$_2$ complex. We use the dipole moment of CO and the total partition function of CO-paraH$_2$ calculated from the now known energy level scheme of the complex (see Fig.~\ref{f:energylevels}). Our calculations result in a value of $\sim 3\times10^{12}$~cm$^{-2}$ for the column density of the complex and a fractional abundance of the CO-paraH$_2$ dimer relative to CO of $\sim 5\times10^{-6}$, assuming the CO column density to be $6\times10^{17}$~cm$^{-2}$ (derived from the $^{13}$C$^{18}$O column density listed in Table~\ref{t:molecules}, and assuming standard $^{12}$C/$^{13}$C and $^{16}$O/$^{18}$O ratios of 60 and 500, respectively.

In the following, we estimate what number density of the CO-H$_2$ molecular complex do we expect under the ISM conditions, and compare it to the new upper limit. All the reaction rates used in the following are generic rates, which have not specifically been measured or calculated, and are taken from the review paper by   \citet{vD14}. The given reactions are the basic types of reactions in space. Following \citet{vD14}, there are two basic processes by which molecular bonds can be formed in the interstellar molecular clouds: radiative association and formation on grain surface with subsequent release to the gas phase. In the radiative association process, the binding energy of a new molecule or molecular complex is carried out through the emission of a photon, and can be described as:
\begin{equation}
\mathrm{H}_2 + \mathrm{CO} \rightarrow \mathrm{CO}\mathrm{-}\mathrm{H}_2 + h\nu
\end{equation}
and proceeds at the rate of a radiative association reaction $k_1\approx10^{-17}$--$10^{-14}$~cm$^3$~s$^{-1}$. For the case of the formation on grain surfaces, a dust particle accommodates the released energy, and the process can be described as
\begin{equation}
\mathrm{H}_2 + \mathrm{CO}\mathrm{-}\mathrm{grain} \rightarrow \mathrm{CO}\mathrm{-}\mathrm{H}_2 + \mathrm{grain}\nonumber
\end{equation}
which proceeds at a rate of $k_2\approx10^{-17}$~cm$^3$~s$^{-1}$.

On the other side, there are three processes for the destruction of the complex: photodissociation, collisional dissociation and neutral-neutral bond rearrangement. The first one can be described by
\begin{equation}
\mathrm{CO}\mathrm{-}\mathrm{H}_2 +  h\nu \rightarrow \mathrm{products}
\end{equation}
with a reaction rate of $k_3\approx10^{-10}$--$10^{-8}$~cm$^3$~s$^{-1}$. The second and third ones can be given by
\begin{equation}
\mathrm{CO}\mathrm{-}\mathrm{H}_2 +  \mathrm{M} \rightarrow \mathrm{products}
\end{equation}
where M being a reaction partner, with rates for collisional dissociation of $k_4\approx10^{-26}$~cm$^3$~s$^{-1}$ and for bond rearrangement of $k_5\approx10^{-11}$--$10^{-9}$~cm$^3$~s$^{-1}$.

We consider a dense condensation with an H$_2$ density of [H$_2$] = $4\times10^{4}$~cm$^{-3}$, the CO density given by [CO] = [CO-grain] = 10$^{-4}$ [H$_2$], and assume [M] = [H$_2$]. Under these conditions, {the formation is dominated by radiative association, while the destruction mainly occurs by the bond rearrangement. As it is stated by \citet{vD14}, collisional dissociation of molecules is only important in regions of very high temperature ($>3000$~K) and density. Thus,} we determine the CO-H$_2$ abundance in the equilibrium as [CO-H$_2$] = ($k_1$[H$_2$][CO])/($k_5$[M]) = $4\times10^{-8}$--$4\times10^{-3}$~cm$^{-3}$ and [CO-H$_2$]/[CO] $\sim 10^{-8}$--$10^{-3}$. The obtained range for a possible abundance of CO-H$_2$ is quite wide. From the comparison of our estimated [CO-H$_2$]/[CO] abundance to the upper detection limit of [CO-H$_2$]/[CO]$\sim5\times10^{-6}$, we can conclude that the complex might be detected by observations with one or two orders higher sensitivity.

\subsection{Molecular inventory in cold regions\label{s:moleculesDisc}}

Table~\ref{t:molecules} and Figure~\ref{f:molecules} reveal a relatively rich chemistry in the cold dense core TMC-1C. Despite the average excitation temperature being of only 7~K, we are able to detect a number of species with 6 or more atoms: CH$_3$CN, CH$_3$OH and CH$_3$CCH. The column densities for these species are in the range $10^{11}$--$10^{13}$~cm$^{-2}$, which results in abundances of $10^{-12}$--$10^{-10}$ assuming a H$_2$ column density of 10$^{22}$~cm$^{-2}$ (e.g.\ \citealt{Schnee2005}). These abundances are about two orders of magnitude lower than the typical abundances found toward more massive hot molecular cores. We have searched for more complex species, such as methyl formate (CH$_3$OCHO) or dimethyl ether (CH$_3$OCH$_3$), but we have not detected them with an upper limit on the column density of about $10^{12}$~cm$^{-2}$, assuming an excitation temperature of 7~K. Similarly to the cold core L1689B studied by \citet{Bac12} we also detect ketene (CH$_2$CO), with a column density of $\sim 3\times10^{12}$~cm$^{-2}$ in complete agreement with the column densities determined for L1689B. In addition to the main isotopologues of the detected species, we also detect transitions of the deuterated counterparts CH$_3$CCD and CH$_2$DOH. The deuteration level is estimated to be about 0.045 for CH$_3$CCH and 0.055 for CH$_3$OH, however, this deuteration fractions should be better constraint with future observations of other transitions and with higher spectral resolution (necessary to resolve the lines). The uncertainty of the column density listed in Table~\ref{t:molecules} does not includes the uncertainty in the linewidth, which can not be measured in our coarse spectral resolution observations. The column densities can differ by 30\% if the linewidth is increased/decreased by 0.1~km~s$^{-1}$, or by 50\% if the variation is 0.2~km~s$^{-1}$. Therefore, the column densities reported in Table~\ref{t:molecules} have to be considered with caution. High-spectral resolution observations are necessary to improve the determination of the excitation temperature and column density. Another source of uncertainty in the column density determination is the excitation temperature: Observations of more transitions for the different molecules are required to better constraint the column density and to search for non-LTE effects.

In general, a number of deuterated compounds have been detected: DCS$^{+}$, HDCS, NH$_2$D, c-C$_3$HD, c-C$_3$D$_2$, CH$_2$DOH and CH$_3$CCD. The deuteration fraction is 0.2 for H$_2$CS, 0.07 for c-C$_2$HD, and about 0.05 for CH$_3$CCH and CH$_3$OH. It is worth noting that the column density measured for c-C$_3$HD and c-C$_3$D$_2$ is in agreement with the recent measurements of \citet{Spez13}.

Finally, in addition to the COMs discussed above, we highlight the detection of some species: (\textit{a}) HCS$^{+}$ has been observed in previous surveys towards Taurus molecular cores (e.g.\ \citealt{Oh98, Kai04}). Here, we present for the first time, a tentative detection of the deuterated counterpart DCS$^{+}$. A detailed study of different deuterated species may help to better understand the routes of deuteration, in particular for those more complex species, and to compare with similar studies conducted in high-mass star forming regions (e.g.\ \citealt{Fontani2011, Fontani2015}); (\textit{b}) Similarly, we report for the first time a tentative detection of HOCO$^{+}$ in this source, for which we determine a column density of $\sim 2\times10^{11}$~cm$^{-2}$; and (\textit{c}) The detection of HCO is common in photon-dominated regions (PDRs; e.g.\ \citealt{schilke2001}), where the chemistry is dominated by the presence of large amounts of far-UV photons. The Taurus molecular cloud is a low-mass star forming complex, and therefore there are no high-mass stars in the region able to produce enough UV photons. In this survey we report the detection of HCO in a cold, dense core, not associated with a PDR, with a column density of $\sim 10^{12}$~cm$^{-2}$. \citet{Bacmann2016} studied HCO in a number of cold prestellar cores, and related its abundance with that of other species such as H$_2$CO, CH$_3$O and finally CH$_3$OH. The authors determine the abundance ratios between the different species to be HCO:H$_2$CO:CH$_3$O:CH$_3$OH $\sim 10:100:1:100$, when the formation of methanol occurs via hydrogenation of CO on cold grain surfaces. The observed abundances of the intermediate species HCO and CH$_3$O suggest they are gas-phase products of the precursors H$_2$CO and CH$_3$OH, respectively. We measure an abundance ratio of HCO:CH$_3$OH $\sim 1:10$ for our cold, dense core (see Table~\ref{t:molecules}), consistent with the results reported by \citet{Bacmann2016}.

\section{Summary\label{s:summary}}

We have used the IRAM\,30m telescope to conduct sensitive observations of a cold, dense core in TMC-1C, with the goal of detecting the CO-H$_2$ van der Waals complex. We have not detected any transition of the CO-paraH$_2$ and CO-orthoH$_2$ compounds with a rms noise level of $\sim 2$~mK for a spectral resolution of 0.7~km~s$^{-1}$. This sets a new strong upper limit for the abundance of the complex: [CO-H$_2$]/[CO]~$\sim 5\times10^{-6}$. We estimate that the expected abundance of the complex, with respect to CO, in the ISM can be $\sim 10^{-8}$--$10^{-3}$, which suggest that more sensitive observations would be required to search for and detect for the first time the CO-H$_2$ complex in the ISM. 

Our sensitive spectral line survey have revealed the detection of 75 different spectral lines coming from 41 different species (including isotopologues). The excitation temperature is $\sim 7$~K, consistent with previous estimates. We detect a number of complex organic molecules such as CH$_3$CN, CH$_3$OH, CH$_3$CCH and deuterated isotopologues. The detection of these species in a cold object is consistent with the similar findings in other objects (e.g.\ L1689B, \citealt{Bac12}). Future studies of these complex species to better constraint the physical parameters, as well as the study of more rare isotopologues, can help to improve the current understanding of the formation of complex species in the cold ISM.

\begin{acknowledgements}
We acknowledge the comments and suggestions of the anonymous referee that helped to improve the manuscript. AP would like to thank Nicolas Billot for his help with the observations and data processing and the IRAM team. This work was supported by Deutsche Forschungsgemeinschaft through grant SFB 956 (subprojects A6, B4 and C3).
\end{acknowledgements}

\bibliographystyle{aa} 
\bibliography{biblioIRAM} 

\end{document}